\documentclass{iau}
\usepackage{graphicx}
\newcommand{\ion}[2]{#1\,{\sc{#2}}}

\title[Accretion and outflow of gas in Markarian 509] 
{Accretion and outflow of gas \\in Markarian 509}

\author[Jelle Kaastra et al.]   
{Jelle Kaastra $^{1,2}$
 \and  Pierre-Olivier Petrucci $^3$
 \and  Massimo Cappi $^4$
 \and  Nahum Arav $^5$
 \and  Ehud Behar $^6$
 \and  Stefano Bianchi $^7$
 \and  Graziella Branduardi-Raymont $^8$
 \and  Elisa Costantini $^1$
 \and  Jacobo Ebrero $^1$
 \and  Jerry Kriss $^{9,10}$
 \and  Missagh Mehdipour $^8$
 \and  Stephane Paltani $^{11}$
 \and  Ciro Pinto $^{1}$
 \and  Gabriele Ponti $^{12}$
 \and  Katrien Steenbrugge $^{13,14}$
 \and  Cor de Vries $^1$}

\affiliation{$^1$ SRON Netherlands Institute for Space Research, 
 Sorbonnelaan 2, 3584 CA Utrecht, The Netherlands \\ email: {\tt j.kaastra@sron.nl } \\[\affilskip]
$^2$Utrecht University, Utrecht, The Netherlands \\[\affilskip]
$^3$UJF-Grenoble 1 / CNRS-INSU, Institut de Plan\'etologie et d'Astrophysique de Grenoble (IPAG) UMR 5274, Grenoble 38041, France \\[\affilskip]
$^4$INAF-IASF Bologna, via Gobetti 101, 40129 Bologna, Italy  \\[\affilskip]
$^5$Department of Physics, Virginia Tech, Blacksburg, VA 24061, USA  \\[\affilskip]
$^6$Department of Physics, Technion-Israel Institute of Technology, 32000 Haifa, Israel  \\[\affilskip]
$^7$Dipartimento di Fisica, Universit\'a degli Studi Roma Tre, via della Vasca Navale 84, \\ 00146 Roma, Italy  \\[\affilskip]
$^8$Mullard Space Science Laboratory, University College London, Holmbury St. Mary, Dorking, Surrey, RH5 6NT, UK  \\[\affilskip]
$^9$Space Telescope Science Institute, 3700 San Martin Drive, Baltimore, MD 21218, USA  \\[\affilskip]
$^{10}$ Department of Physics and Astronomy, The Johns Hopkins University, Baltimore, \\ MD 21218, USA  \\[\affilskip]
$^{11}$ ISDC Data Centre for Astrophysics, Astronomical Observatory of the University of Geneva, 16, ch. d'Ecogia, 1290 Versoix,
Switzerland  \\[\affilskip]
$^{12}$ Max-Planck Institut f\"ur Extraterrestrische Physik, Garching, Germany  \\[\affilskip]
$^{13}$ Instituto de Astronom\'ia, Universidad Cat\'olica del Norte, Avenida Angamos 0610, Casilla 1280, Antofagasta, Chile  \\[\affilskip]
$^{14}$ Department of Physics, University of Oxford, Keble Road, Oxford OX1 3RH, UK}

\pubyear{2012}
\volume{290}  
\jname{Feeding compact objects: Accretion on all scales}
\editors{C.M. Zhang, T. Belloni, M. M\'endez \& S.N. Zhang, eds.}

\begin{document}

\maketitle

\begin{abstract}
A major uncertainty in models for photoionised outflows in AGN is the distance
of the gas to the central black hole. We present the results of a massive
multiwavelength monitoring campaign on the bright Seyfert 1 galaxy Mrk~509 to
constrain the location of the outflow components dominating the soft X-ray band.

Mrk~509 was monitored by XMM-Newton, Integral, Chandra, HST/COS and Swift in
2009. We have studied the response of the photoionised gas to the changes in the
ionising flux produced by the central regions. We were able to put tight
constraints on the variability of the absorbers from day to year time scales.
This allowed us to develop a model for the time-dependent photoionisation in
this source. 

We find that the more highly ionised gas producing most X-ray line opacity is at
least 5~pc away from the core; upper limits to the distance of various absorbing
components range between 20~pc up to a few kpc. The more lowly ionised gas
producing most UV line opacity is at least 100~pc away from the nucleus. 

These results point to an origin of the dominant, slow ($v$$<$1000 km\,s$^{-1}$) outflow
components in the NLR or torus-region of Mrk~509. We find that while the
kinetic luminosity of the outflow is small, the mass carried away is likely
larger than the 0.5~Solar mass per year accreting onto the black hole. 

We also determined the chemical composition of the outflow as well as valuable
constraints on the different emission regions. We find for instance that the
resolved component of the Fe-K line originates from a region 40--1000
gravitational radii from the black hole, and that the soft excess is produced by
Comptonisation in a warm (0.2--1 keV), optically thick ($\tau\sim $10--20) corona near
the inner part of the disk.

\keywords{Active Galactic Nuclei, X-rays, outflows}
\end{abstract}

\firstsection 
\section{Introduction}

Active Galactic Nuclei (AGN) are powered by accretion onto a super-massive black
hole. In addition to the accretion process, however, it has been found that many
AGN contain strong outflows. The total mass loss associated with this outflow
may even be larger than the mass that finally reaches the black hole, and the
kinetic energy associated with the outflow may be large enough to regulate the
growth of the black hole and tightly connected to it, the growth of the host
galaxy.  

Different origins for these outflows or winds have been proposed, like accretion
disk winds, thermal evaporation from the torus, or extended ionisation cones.
These models all imply very different effects for the impact of the outflow on
the environment, the feedback process. First, outflows launched close to the
black hole (like accretion disk winds) have to overcome a larger gravitational
potential to escape, hence should have much larger outflow velocities $v$, from
thousands of km\,s$^{-1}$ to significant fractions of the speed of light.
Furthermore, outflows at larger distances $r$ from the black hole (like e.g.
torus winds) have a larger impact on the surroundings, because we have for the
mass loss rate $\dot M$: \begin{equation} \dot M / \Omega = N_{\rm H} m_p r v,
\label{eqn:mdot} \end{equation} with $\Omega$ the solid angle subtended by the
outflow, $m_p$ the proton mass and $N_{\rm H}$ the total column density. The
kinetic luminosity $L_K$ is simply given by $\frac{1}{2}\dot M v^2$. 

While column densities and outflow velocities can be obtained directly from the
spectrum, the distance should be obtained by other means. Given the ionisation
parameter $\xi=L/nr^2$ with $L$ the ionising luminosity and $n$ the hydrogen
density, a known value for $n$ immediately gives $r$, because $L$ and $\xi$ are
known from observations. Because the recombination time scale of the plasma
scales with $n^{-1}$, the delay time of the ionisation state of the plasma with
respect to variations of the continuum luminosity immediately yields the density
and therefore the distance of the absorber. This is the principle that forms the
basis for our large monitoring campaign on Mrk~509, a bright and luminous
Seyfert~1 galaxy at $z=0.034$.

\section{The monitoring campaign on Mrk~509}

We have observed Mrk~509 intensively during a 100-day campaign in 2009
\cite[(Kaastra et al. 2011a)]{kaastra2011a}. The core of the campaign consisted
of 10$\times$60~ks observations with XMM-Newton, spaced 4 days, coincident with
ten INTEGRAL observations of 120~ks each. This was followed by a 180~ks
observation with the Chandra LETGS, simultaneous with 10 orbits HST/COS. It was
preceded with Swift monitoring (UV, X-ray) and supplemented with ground-based
observations (WHT, Pairitel). Up to now, 12 refereed papers have been published
on this campaign and more are under preparation.

The lightcurve of Mrk~509 \cite[(Kaastra et al. 2011a)]{kaastra2011a} showed the
expected variability with a large ($\sim$60\% increase) outburst in the middle
of our campaign. Our first step was to determine the time-averaged spectrum.
Because of the high statistical quality of the data, special analysis tools and
refinements to the calibration needed to be developed, in particular for the
RGS instruments of XMM-Newton \cite[(Kaastra et al. 2011b)]{kaastra2011b}, but
also for the EPIC camera \cite[(Ponti et al. 2012)]{ponti2012} and the Optical
Monitor \cite[(Mehdipour et al. 2011)]{mehdipour2011} of XMM-Newton and the HST/COS
spectrometer \cite[(Kriss et al. 2011)]{kriss2011}.

The time-averaged spectrum showed five different ionisation components producing
significant absorption lines in the RGS spectrum \cite[(Detmers et al.
2011)]{detmers2011}. These results were confirmed by the LETGS spectra
\cite[(Ebrero et al. 2011)]{ebrero2011}. The quality of the data was high enough
to prove for the first time that two of the ionisation components, C and D, are
discrete structures with a FWHM of less than 35 and 80\%, respectively. However,
these components are {\sl not} in pressure equilibrium. From component A to E,
the gas pressure drops by two orders of magnitude. For a
proper modeling of the ionisation structure the spectral energy distribution
(SED) of the source needs to be known accurately, and thanks to our broad-band
coverage from optical to hard X-ray wavelengths we were able to do this
modeling.

The next step was to constrain the distance $r$ of the absorbers, by measuring
the delayed response of the absorbers with respect to the continuum variations.
To do this properly, we developed a time-dependent photo-ionisation model.
Starting with equilibrium, we followed the spectral variations over time and
calculated the ion concentrations for different gas densities $n$. Comparing the
predicted changes in absorber transmission for each of these models with the
observed spectra, the distances could be constrained \cite[(Kaastra et al.
2012)]{kaastra2012}. For components C--E no significant variability during our
campaign was found, leading to lower limits of their distance. However,
component D showed evidence for variability on long time scales, by comparison
with archival spectra taken in 2000 and 2001, yielding a tight upper limit to
its distance. For components C and E upper limits were derived from the
requirement that $r<L/N_{\rm H}\xi$. Our measurements were not sensitive enough
to constrain variability for components A and B, but  \ion{O}{iii} is present in
component A, and through direct imaging of its forbidden 5007~\AA\ line its
distance has been determined to be $\sim$3~kpc \cite[(Phillips et al.
1983)]{phillips1983}. Furthermore, from variability constraints of the UV line
velocity components, corresponding to ionisation components A and B, \cite[Arav
et al. (2012)]{arav2012} derived lower limits of $\sim$200~pc for most of them. 

\begin{figure}[htbp]
\begin{center}
 \includegraphics[width=3.4in,angle=-90]{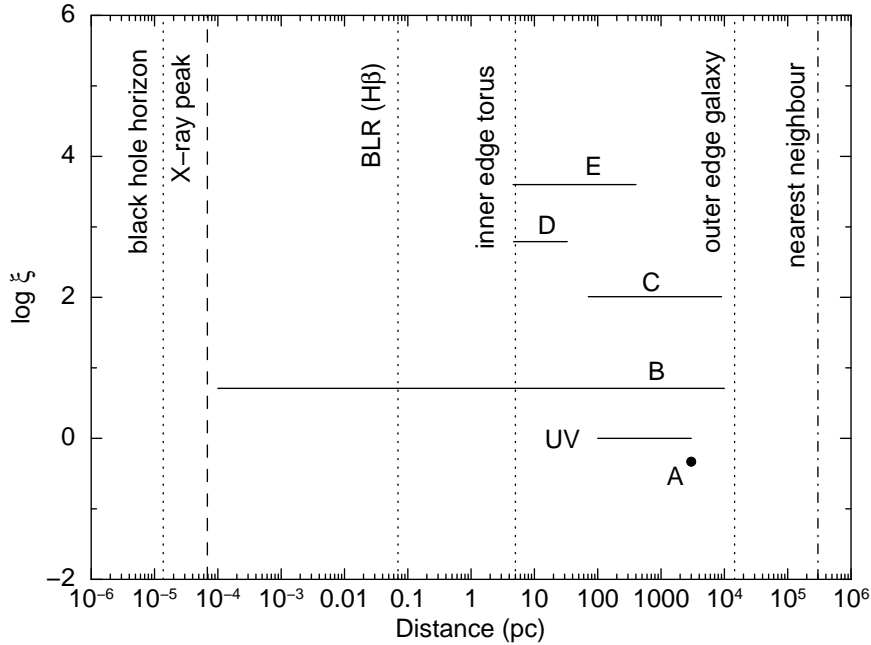}
\caption{Distances of the absorption components in Mrk~509. Horizontal bars
indicate the allowed ranges for each of the individual absorption components.
The x-axis gives distances from the black hole, the y-axis the ionisation
parameter $\xi$ for each component A--E. The constraint from the UV lines, a
mixture of ionisation components A and B, is also shown. Vertically dashed lines
indicate some radii of interest.}
   \label{fig1}
\end{center}
\end{figure}

Fig.~\ref{fig1} summarises the distance limits. The dominant X-ray absorption
components C--E originate from a region outside the inner torus radius. Thus,
these outflow components are more likely explained by torus winds rather than
accretion disk winds.

Using several arguments, the total mass loss of the wind is between 0.26 and
2100 solar masses per year. This is large compared to the accretion rate of
about 0.5 solar mass per year. However, due to the limited outflow velocities of
the wind (up to several 100 km\,s$^{-1}$, but not relativistic), the kinetic
luminosity of the outflow is not very large.

Our spectra also yielded the metal abundances of the outflow.  \cite[Steenbrugge
et al. (2011)]{steenbrugge2011} measured the abundances of C, N, Ne, Mg, Si, S,
Ca and Fe relative to iron with accuracies down to 8\% for some elements. The
ratio's agree fairly well with solar abundances, although we could not
yet constrain the absolute abundances. We will determine those later using new
HST/COS spectra covering better the Lyman series of hydrogen.

The Galactic foreground absorption was studied by \cite[Pinto et al.
(2012)]{pinto2012}. The UV spectrum shows seven discrete absorption structures.
Some of these are associated to high-velocity clouds, others correspond to
absorption within the spiral arms of our Galaxy. A detailed assessment of the
role of dust and gas at various temperatures could be made.

The UV lightcurve shows a distinct outburst near the centre of our monitoring
campaign. The soft X-ray flux correlates very well with these variations in the
UV flux, while the hard X-ray flux does not correlate at all with the other
bands. This and the spectral shape led \cite[Mehdipour et al.
(2011)]{mehdipour2011} to conclude that the soft X-ray excess in Mrk~509 is not
caused by blurred reflection, but by comptonisation of the soft UV photons of
the disk. Further detailed modeling by \cite[Petrucci et al.
(2012)]{petrucci2012} strenghtened this picture. The scattering medium has a
large optical depth of $\sim$15, at a temperature of $\sim$1~keV.

Finally, \cite[Ponti et al. (2012)]{ponti2012} studied the Fe-K emission line,
showing that the neutral component consists of a narrow, constant component,
possibly originating from material between the outer broad-line region and
the torus, and a broad variable
component, originating from the inner broad line region.




\begin{thebibliography}{}

\bibitem[Arav et al. (2012)]{arav2012}
{Arav, N., Edmonds, D., Borguet, B., et al.} 2012, 
\textit{A\&A}, 544, A33 

\bibitem[Detmers et al. (2011)]{detmers2011}
{Detmers, R.G., Kaastra, J.S., Steenbrugge, K.C., et al.} 2011, 
\textit{A\&A}, 534, A38 

\bibitem[Ebrero et al. (2011)]{ebrero2011}
{Ebrero, J., Kriss, G.A., Kaastra, J.S., et al.} 2011, 
\textit{A\&A}, 534, A40

\bibitem[Kaastra et al. (2011a)]{kaastra2011a}
{Kaastra, J.S., Petrucci, P.-O., Cappi, M., et al.} 2011a, 
\textit{A\&A}, 534, A36

\bibitem[Kaastra et al. (2011b)]{kaastra2011b}
{Kaastra, J.S., de Vries, C.P., Steenbrugge, K.C., et al.} 2011b, 
\textit{A\&A}, 534, A37

\bibitem[Kaastra et al. (2012)]{kaastra2012}
{Kaastra, J.S., Detmers, R.G., Mehdipour, M., et al.} 2012, 
\textit{A\&A}, 539, A117 

\bibitem[Kriss et al. (2011)]{kriss2011}
{Kriss, G.A., Arav, N., Kaastra, J.S.,  et al.} 2011, 
\textit{A\&A}, 534, A41   

\bibitem[Mehdipour et al. (2011)]{mehdipour2011}
{Mehdipour, M., Branduardi-Raymont, G., Kaastra, J.S., et al.} 2011, 
\textit{A\&A}, 534, A39

\bibitem[Petrucci et al. (2012)]{petrucci2012}
{Petrucci, P.-O., Paltani, S., Malzac, J., et al.} 2012, 
\textit{A\&A}, in press (arXiv1209.6438)

\bibitem[Phillips et al. (1983)]{phillips1983}
{Phillips, M.M., Bladwin, J.A., Atwood, B. \& Carswell, R.F.} 1983, 
\textit{ApJ}, 274, 558
 
 \bibitem[Pinto et al. (2012)]{pinto2012}
{Pinto, C., Kriss, G.A., Kaastra, J.S., et al.} 2012, 
\textit{A\&A}, 541, A147 

\bibitem[Ponti et al. (2012)]{ponti2012}
{Ponti, G., Cappi, M., Costantini, E., et al.} 2012, 
\textit{A\&A}, in press (arXiv1207.0831)
 
\bibitem[Steenbrugge et al. (2011)]{steenbrugge2011}
{Steenbrugge, K.C., Kaastra, J.S., Detmers, R.G., et al.} 2011, 
\textit{A\&A}, 534, A42

\end{thebibliography}
\end{document}